# The Role of Oxygen Defects on the Electro-Chemo-Mechanical Properties of Highly Defective Gadolinium Doped Ceria

Ahsanul Kabir[1,2], Jin Kyu Han[1], Benoit Merle[2], and Vincenzo Esposito[1*]

[1]Department of Energy Conversion and Storage, Technical University of Denmark, Frederiksborgvej 399, Roskilde 4000, Denmark

[2]Materials Science & Engineering I, University Erlangen-Nürnberg (FAU), Cauerstr. 3, 91058 Erlangen, Germany

*Corresponding Authors: E-mail: ahsk@dtu.dk, vies@dtu.dk

## Abstract

In light of the recent discovery of giant electrostriction in defective fluorites, here we investigate the interplay between mechanical, electrochemical and electromechanical properties of oxygen defective ceria compositions ($Ce_{1-x}Gd_xO_{2-\delta}$) as the effect of Gd-doping (x = 0.05−0.3) at low temperatures. Highly dense polycrystalline ceramics are prepared as micron-size grains with a minimized grain boundary extent. Electrochemical ionic migration reveals that doping controls the configuration of oxygen vacancies in the samples. Unexpectedly, we observe that electromechanical activity is depends on oxygen vacancy configuration rather than on its nominal concentration. The primary creep at room temperature indicates a declining viscoelastic trend with increasing oxygen defects.

Keywords: Defects, Blocking Barriers, Ionic conductivity, Creep, Electrostriction,

1. Introduction

Electromechanically active materials have shown growing interests in a wide range of sensing and actuating applications including consumer electronics, ultrasound transducers, sonar, *etc.* [1]. Electrostriction is the second-order electromechanical coupling developed in all insulators, with a strong dependency on material´s dielectric permittivity (ε) and elastic compliance (S). A recent investigation illustrates that specific orientated thin films of highly defective Gd-doped cerium oxides ($Ce_{0.8}Gd_{0.2}O_{1.9}$) display a non-classical giant electrostriction response with $M_e$ ≈ 6.5 x $10^{-18}$ $m^2/V^2$ at 0.1 Hz, which is at least two orders of magnitude larger than classical prediction [2]. Moreover, the polycrystalline bulk form of GDC and another similar oxide ($Bi_2O_3$) also exhibit similar results of even higher magnitudes [3][4][5]. The atomistic mechanism of such mechanism is attributed to the presence of electroactive elastic dipoles ($Ce_{Ce}$–$O_O$) that changes their bond length under an external electric field [2][6]. Despite experimental shreds of evidence indicating that oxygen vacancies-cation complexes play a key role in the electromechanical properties, the concentration of the defects is generally considered the primary parameter in controlling the electrostriction in defective fluorites [4][7]. Similar to electrostriction, $Ce_{Ce}$–$O_O$ dipole rearranges under mechanical load, exhibiting an unusual primary creep behavior at room temperature in nanoindentation measurement [4]. Moreover, oxygen defects can take different configurations in the fluorites, especially in polycrystalline materials with different degrees of disorder and composition inside the grains and at grain boundaries [8]. These effects are observed for oxygen ionic conductivity [8] as well as for cation diffusion at high temperatures [9][10][11], where both electrostatic and steric effects between dopants and oxygen defects can control the energetic barrier of such processes [12]. In this work, we demonstrate the role of oxygen vacancies on the electrical, mechanical and electromechanical properties of Gd-doped ceria, highlighting their concentration/configuration dependency on each effect.

## 2. Experimental Procedure

The nanometric gadolinium doped ceria (GDC) powders with composition $Ce_{1-x}Gd_xO_{2-\delta}$ where x = 0.05−0.3 were synthesized by the co-precipitation method, as described elsewhere [5][8]. The powders were uniaxially cold-pressed at 200 MPa and subsequently sintered at 1450 °C in air for 10 h. The density of the pellets was measured by the Archimedes method in deionized water. The crystallographic phase purity was analyzed by the X-ray diffraction (XRD) technique (Bruker D8, Germany). The microstructure was characterized by a high-resolution scanning electron microscope (SEM; Zeiss Merlin, Germany). The electrical conductivity was measured by electrochemical impedance spectroscopy (EIS) Solarton 1260 (UK) in a temperature range of 300–450 °C for a frequency distribution of 0.01–$10^7$ Hz in air. The electromechanical property is examined in a nano-vibration analyzer with a single-beam laser interferometer (SIOS, Germany). The mechanical properties of the samples were investigated by nanoindentation technique (KLA G200, USA), using Berkovich indenter at room temperature. The creep measurements were based on a typical trapezoidal load-hold-unload system with a loading/unloading rate of 15 mN/s (fast) or 1.5 mN/s (slow) and holding at load 150 mN or depth of 1000 nm for 20 s duration (see **Fig. 2a**). During holding, the creep relaxation of the material was recorded as a progressive increase in displacement. The elastic modulus and nano-hardness of the material were determined by the Oliver-Pharr analysis [13]. For the local strain rate sensitivity measurement, the strain rate was varied between 0.001 $s^{-1}$ and 0.1 $s^{-1}$ and evaluated from the resulting jump in hardness [14][15].

## 3. Results and Discussion

The XRD patterns reveal characteristic cubic fluorite structure (Fm−3m), confirming a formation of a single-phase solid solution (see **Fig. S1.a**). A highly dense microstructure was observed, which agrees with the results of experimental density (> 95% of theoretical density).

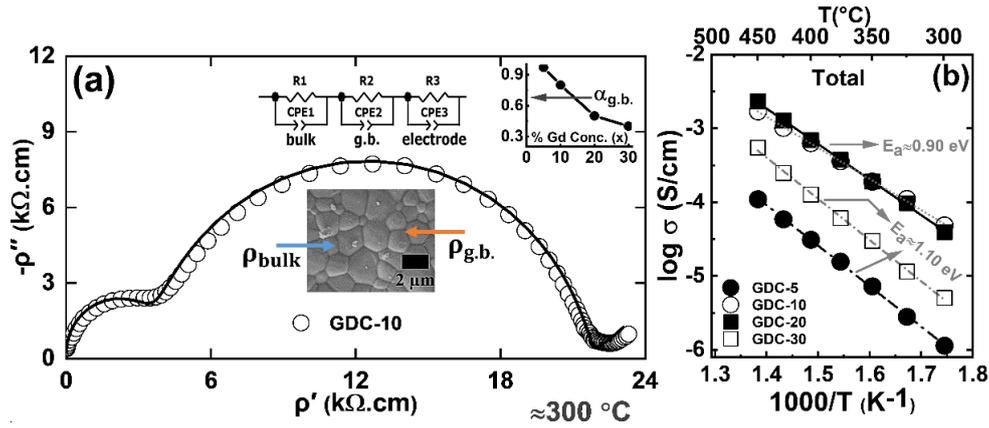

**Figure 1:** (a) Illustration of a characteristic impedance spectrum of GDC-10 sample, measured at 300 °C in air with silver (Ag) electrode. (b) The Arrhenius plot for the estimation of total electrical conductivities of the sintered GDC pellets.

The thermal (binding) energy and mobility of oxygen vacancies is characterized by complex impedance spectroscopy and **Fig. 1a** demonstrates the typical impedance plot (Nyquist formalism) at 300 °C in ambient air. The semicircles at the high and intermediate frequency refer to bulk and grain boundary impedance, respectively. The plot highlights the arising of ion blocking factors, usually associated with the disorder at the grain boundary (g.b.) [8][16]. The ion blocking effect is characterized by the grain boundary blocking factor ($\alpha_{g.b.}$) where $\alpha_{g.b.} = \frac{R_{gb}}{R_{bulk} + R_{g.b.}}$ [8], shows to decrease with increasing Gd concentration. The temperature dependence total electrical conductivity ($\sigma$) of the sample is shown in **Fig. 1.b** in an Arrhenius plot. The result illustrates that GDC-10 and GDC-20 samples possess maximum conductivity which is half-one order higher than counterparts are, associating to low activation energy value (≈0.9 eV) in the materials. **Fig. S2** describes how the total conductivity is controlled by the g.b. blocking effects. For instance, the GDC-5 sample has low bulk migration energy but a high blocking factor, leading to lowermost ionic conductivity.

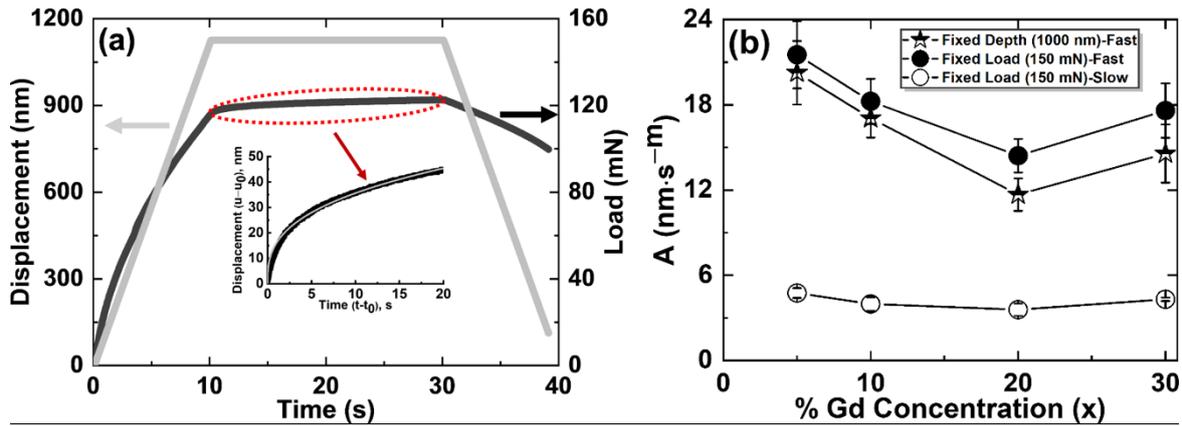

**Figure 2:** (a) The representative load–time and displacement–time curve of e GDC-10 sample in the nanoindentation fast loading approach. The inset shows a non-linear displacement–time plot in the holding segment, indicating a primary creep. (b) The creep constant (*A*) as a function of Gd-doping in ceria compositions under fixed depth (fast) and fixed load (fast/slow) mode.

All investigated samples display noticeable creep behavior during the hold section of the measurement under fast loading. The holding time dependence of the displacement shows a non-linear relation, following an empirical formula $(u - u_0) = A\,(t - t_0)^m$ where $u_0$ is the initial displacement at time $t_0$ at the holding segment, *A* is defined as creep constant and *m* the fitting exponent. The creep constant *A* is shown in **Fig. 2b** as a function of Gd-concentration. *A* decreases linearly for 0.05 < Gd < 0.2, in agreement with previous reports, suggesting that the increased interaction between cations and oxygen vacancies is responsible for the lesser creep and the increasing hardness [17][18]. These results are consistent with the impedance, where high dopant content reduces the conductivity at the bulk (see **Fig. S2**). The augmented value for GDC-30 could be attributed to the possible local double fluorite structure of the material [19]. As expected, the creep constant is considerably higher for a fast pre-loading than for the slow mode. Its value hardly depends on whether the creep segment starts after reaching 150 mN or 1000 nm. The exponent *m* ranges between 0.3–0.6 nm · s$^{-m}$ and slightly decreases with Gd-content, similar to the strain rate sensitivity (see **Fig. S4 and S5**).

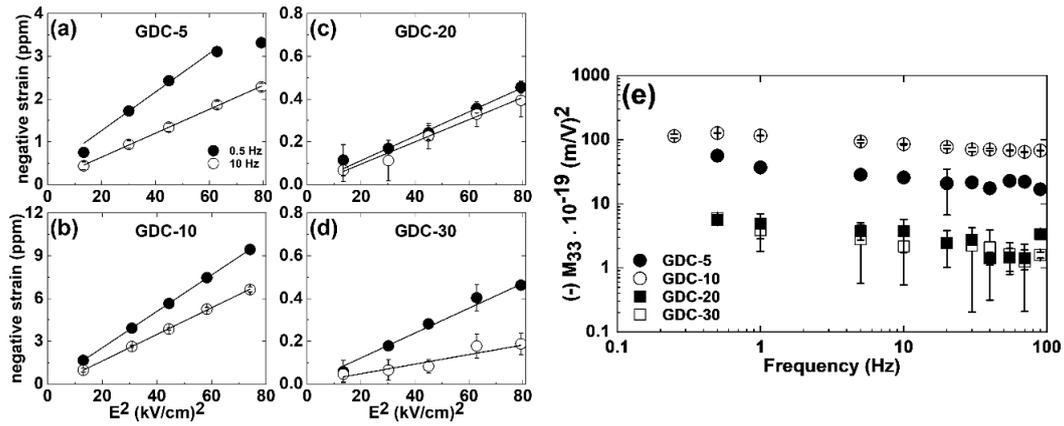

**Figure 3:** (a)-(d) The electrostrictive negative strain with the response to the external electric field at frequencies 0.5 and 10 Hz for GDC samples (Gd = 0.05–0.3), electrode material: Gold (Au). (e) The electrostrictive strain coefficient ($M_{33}$) as a function of frequencies ranging from 0.15–100 Hz.

**Fig. 3** shows that all samples strained negatively at the second harmonic of the applied electric field, verifying the electrostriction behavior of ceria, as reported in [2][4]. A large strain is noticed for low-doped samples that contain large blocking barrier effect ($\alpha_{g.b.} > 0.8$) as well as low oxygen vacancy ordering. As noticed, the strain value is higher at 0.5 Hz than 10 Hz for all investigated samples. Consequently, a gradual declination of the electrostrictive strain coefficient ($M_{33}$) occurs as the applied frequencies increases. Moreover, $M_{33}$ still has an order of $\approx 10^{-18}$ (m/V)$^2$ at higher frequency *e.g.* 50 Hz for GDC-5 and GDC-10 samples that is still one order higher than classical model estimation.

## 4. Conclusion

Highly dense GDC ceramics were successfully produced via conventional sintering with micron-size grains. These samples develop a different blocking effect that scaled-down with Gd concentration, affecting the total ionic conduction as well as electrostriction. They also display non-linear creep properties at room temperature with a strong dependence on nominal Gd content.

## 5. Acknowledgments

This research was supported by DFF-Research project grants (GIANT-E, 48293) June 2016, partially by European H2020-FETOPEN-2016-2017 project BioWings (# 801267) and the